\documentclass[preprint]{ptephy_v1}

\preprintnumber{XXXX-XXXX} 
\usepackage{hyperref}

\usepackage{amsmath} 
\usepackage{amsthm} 
\usepackage{hyperref} 
\usepackage{graphics} 
\usepackage{algorithmic} 
\usepackage{url} 
\usepackage{caption}
\usepackage{subcaption}
\usepackage[table]{xcolor}
\usepackage[section]{placeins}
\usepackage{lineno}
\usepackage[symbol]{footmisc}

\begin{document}

\title{Electromagnetic deflection effects in the integrated luminosity measurement at the CEPC}


\author[1]{Ivan Smiljanić}

\author[1]{Ivanka Božović}
\affil{Vinca Institute of Nuclear Sciences - National Institute of the Republic of
Serbia, University of Belgrade, M. Petrovica Alasa 12-14, Belgrade, Serbia
\email{ibozovic@vin.bg.ac.rs}}

\author[1]{Ivana Vidaković} 

\author[1]{Nataša Vukašinović} 

\author[1]{Goran Kačarević \footnote[2]{left}}


\begin{abstract}%
In order to ensure measurement of the integrated luminosity with a relative precision of $\mathrm{10^{-4}}$ at the $\mathrm{Z^{0}}$ pole at CEPC, numerous systematic effects have to be quantified and, if possible, corrected for. Here we discuss collective impact of electromagnetic bunch fields on the initial state electrons and positrons (EMD1) as well as on the Bhabha scattering final states (EMD2). Both effects change four-momenta of the final state particles, leading to modification of the Bhabha count in the luminometer. These effects are quantified in simulation, together with their stability with respect to the beam parameters variations. Possible correction methods based on experimental measurements with the CEPC detector are discussed on a conceptual level.
\end{abstract}


\maketitle

\section{Introduction}
\label{sec:intro}

    In general, beam bunches at high energy $e^+ e^-$  colliders are densely packed, in order to ensure sufficient instantaneous luminosity and consequently sufficient statistics of processes of interest for realization of physics programs with these machines. In this respect, the proposed beam parameters of the CEPC Conceptual Design Report \cite{CEPC_CDR} have been updated from $\mathrm{8 \cdot 10^{10}}$ to $\mathrm{14 \cdot 10^{10}}$ particles per bunch at the $\mathrm{Z^{0}}$, resulting in even stronger electromagnetic fields \cite{CEPC_postCDR}. These fields will impact the initial states in the bunches of opposite charges, including the effect of the intrinsic bunch field. The collective effect of electromagnetic fields of initial states, labeled as EMD1, is discussed in Section \ref{sec:sec3}, with the CEPC post-CDR beams \cite{CEPC_postCDR}. Of course, electromagnetic deflection of the initial state is not the only effect of collective electromagnetic fields that also provoke beamstrahlung and beam disruption and are basically changing crossing angle geometry and energy-angle correlations of colliding particles. In Section \ref{sec:sec3} we quantify these changes, primarily of the crossing angle and transverse momenta of initial states. It is clear that a change of four-momentum of the system of colliding particles will affect as well the final state electrons and positrons produced in the small-angle Bhabha scattering (SABS), modifying their count in the luminometer compared to the one in the absence of the effect. Once the crossing angle is measured, the effect-driven correction to the integrated luminosity measurement ($\mathrm{\mathcal{L}_{int}}$) can be introduced.

    In addition, electron and positron produced in SABS will be deflected by the fields of opposite-charged bunches (effect labeled as EMD2), focusing them towards the z-axis of the laboratory system of reference. This will result of the loss of count at the luminometer's inner edge and the consequent uncertainty in the integrated luminosity measurement. The EMD2 focusing effect is typically ranging from a few tens of microradians at circular colliders \cite{FCCee} up to a few hundred of microradians at linear colliders \cite{Madison1}. Estimate of EMD2 at the CEPC $\mathrm{Z^0}$ pole is given for the first time in this paper and discussed in detail in Section \ref{sec:sec4}, including the dependence of the effect size on variations of the beam parameters.

    These effects have not yet been experimentally measured. However, they can be estimated from simulation, as it has been done in \cite{FCCee} and \cite{Madison2} for FCCee and ILC, respectively. Although \cite{FCCee} and \cite{Madison2} share the same common tool with this study - GuineaPig C++ software \cite{GuineaPig} to simulate beam-beam interactions, they rather quantify EMD effects in (\cite{FCCee}), or develop a dedicated method to correct the SABS count for EMD2 in (\cite{Madison2}). In this paper we use observables provided by GuineaPig to estimate these effects, compare them to the GuineaPig standard output and to the estimates from other experiments \cite{FCCee}, proving the intrinsic consistency in description and understanding of these effects. In Section \ref{sec:sec5}, at the conceptual level we review possibilities to correct for EMD1 effect on the basis of measurements with the CEPC detector, while dedicated EMD2 correction method study is ongoing by the same authors. As well, additional radiative effects like initial state radiation and beamstrahlung are quantified and discussed in the same section.

\section{Very forward region at CEPC}
\label{sec:sec2}

As described in \cite{CEPC_CDR}, we assume that the machine-detector-interface region of CEPC will be placed in a 118 mrad conus inside the detector, with the luminometer positioned at 95 cm distance from the interaction point, covering the polar angles from 30 mrad to 105 mrad. Fiducial volume (FV) of the luminometer is assumed to be from 53 mrad to 79 mrad,  where the energy resolution will be constant due to shower containment.  The current proposal for luminometer is a LYSO-crystals based luminometer \cite{detector_preTDR}. However we do not discuss technological realization of the luminometer, relying only on the definition of the phase space for the Bhabha count (luminometer fiducial volume). Throughout the paper EMD effects are discussed assuming head-on collision geometry as if the luminometer’s halves would be positioned at the outgoing beams (s-frame), that is 16.5 mrad with respect to the z-axis in the laboratory frame since the CEPC crossing angle is 33 mrad \cite{CEPC_CDR}. GuineaPig software does not inherently simulate detailed RF crab cavity dynamics, so obtained results are comparable to \cite{FCCee}. However, crab-crossing mostly redistributes where the interaction occurs along the longitudinal direction, affecting the electromagnetic deflection of the final state as a sub-leading effect at the level of a few microradians ($\mathrm{\lesssim}$ 10\% of the nominal EMD2 impact) with the resulting impact on the luminosity correction expected to be well below other dominant uncertainties. Also, luminometer resolution to measure polar angle of electrons and positrons scattered in SABS is not exactly considered, since a Si-tracking layer is foreseen to be placed in front of the luminometer, providing polar angle resolution of order of $\mathrm{\mu}$rad \cite{detector_preTDR}. We have assumed Gaussian beams with the beam parameters given in Table\ref{table1} \cite{CEPC_postCDR}. Impact of non-Gaussian beam profile is qualitatively discussed in Section \ref{sec:sec5}. Typical sample size is $\mathrm{10^5}$ Bhabha events\footnote{This limitation in sample sizes comes primarily from the Guinea Pig processing time, even with jobs splitted and processed on Grid.}, corresponding to less than one minute of data collected with the instantaneous luminosity $\mathrm{\mathcal{L}=1.15 \cdot 10^{35} \: cm^{-2}s^{-1}}$ at the $\mathrm{Z^0}$ resonance. We have assumed the beam energy spread being 0.08\% of the nominal beam energy at the $\mathrm{Z^0}$ pole. Also, we introduced radiative effects like initial and final state radiation, as well as beam energy loss due to beamstrahlung and quantified their impact on electromagnetic deflection of initial and final states in Sections \ref{sec:sec3} and \ref{sec:sec5}, respectively.

\begin{table}[!h]
\caption{CEPC accelerator parameters (number of particles per bunches N, $\beta$ function at the interaction point and longitudinal and transverse bunch sizes $\sigma$) assumed in the study.}
\label{table1}
\centering
\begin{tabular}{|c|c|c|c|c|c|}
\hline
N ($10^{10}$) & $\beta^{*}_{x}$ (m) & $\beta^{*}_{y}$ (mm) & $\sigma_{x}$ ($\mathrm{\mu m}$) & $\sigma_{y}$ ($\mathrm{\mu m}$) & $\sigma_{z}$ (mm)\\ 
\hline
14 & 0.13 &  0.9 & 6.0 & 0.036 & 8.7\\
\hline
\end{tabular}
\end{table}

\section{Electromagnetic deflection of the initial state}
\label{sec:sec3}

\begin{figure}[!h]
\centering\includegraphics[width=.6\textwidth]{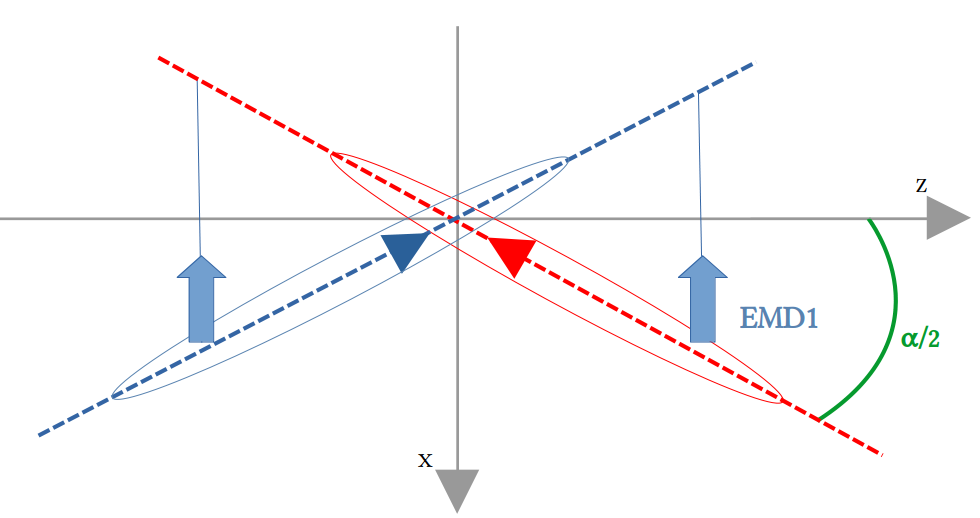}
\caption{Illustration of the EMD1 effect on initial state in the laboratory frame. Colliding bunches of electrons (positrons) are indicated in blue (red).}
\label{fig_1}
\end{figure}

Electromagnetic interaction of initial electron and positron with the bunches of opposite charges is illustrated in Fig. \ref{fig_1}. Initial particles will be attracted toward z-axis in the laboratory frame, resulting in the effective reduction of the crossing-angle $\alpha$. As a result, four-momenta of the final state particles produced in SABS will be modified and consequently their count in the luminometer will change. If uncorrected, this effect will produce a bias in integrated luminosity measurement.

From the analytical point of view, initial-state EM deflection is an event-by-event beam-beam-induced rotation of the instantaneous collision frame, producing angular smearing of the observed Bhabha distribution rather than a dominant first-order polar-angle bias. Each incoming particle traverses the collective field of the opposite bunch and it is subjected to the Lorentz force:

\begin{equation}\label{1.1}
\frac{d\overrightarrow{p}}{dt}=q(\overrightarrow{E}+\overrightarrow{v} \times \overrightarrow{B}),
\end{equation}
where, for ultra-relativistic bunches force is predominantly transverse:

\begin{equation}\label{1.2}
\Delta p_z \ll \Delta p_\perp.
\end{equation}
The incoming particles receive a kick in  the transverse plane:

\begin{equation}\label{1.3}
\Delta \overrightarrow{p}_\perp=q \int{(\overrightarrow{E}_\perp+\overrightarrow{v} \times \overrightarrow{B})_\perp dt},
\end{equation}
corresponding to the effective deflection angle $\mathrm{\Delta \alpha}$:

\begin{equation}\label{1.4}
\Delta \alpha = \frac{|\Delta \overrightarrow{p}_\perp|}{p}.
\end{equation}
Effective deflection angle depends on the bunch parameters (transverse $\mathrm{\sigma_x}$ and longitudinal size $\mathrm{\sigma_z}$, and bunch population N) as:

\begin{equation}\label{1.5}
\Delta \alpha \sim \frac{N}{\sigma_x \cdot \sqrt{(1+\phi_{P}^2})},
\end{equation}
where $\mathrm{\phi_{P} = \sigma_z \alpha / 2\sigma_x}$ is the Piwinski angle. From equation \eqref{1.5} some dependencies can be extrapolated, like smaller the transverse (or longitudinal) bunch sizes or the more populated bunches – deflection of initial states is larger.

To quantify this effect, including collective field of particle’s own bunch with respect to its position in the bunch, we use the GuineaPig C++ V.1.2.2 software, generating around $\mathrm{6 \cdot 10^5}$ electron-positron interacting pairs. GunieaPig is essentially a beam–beam interaction simulator, so it models how the bunch fields distort the incoming particles’ energies, directions, and overlap conditions. GuineaPig does not treat bunches as rigid objects. Typically, bunches are sliced longitudinally, electromagnetic fields are computed slice-by-slice and particles are propagated through evolving collective fields. As a result, the initial state becomes dynamically modified during collision itself.

\begin{figure}[!h]
\centering\includegraphics[width=.6\textwidth]{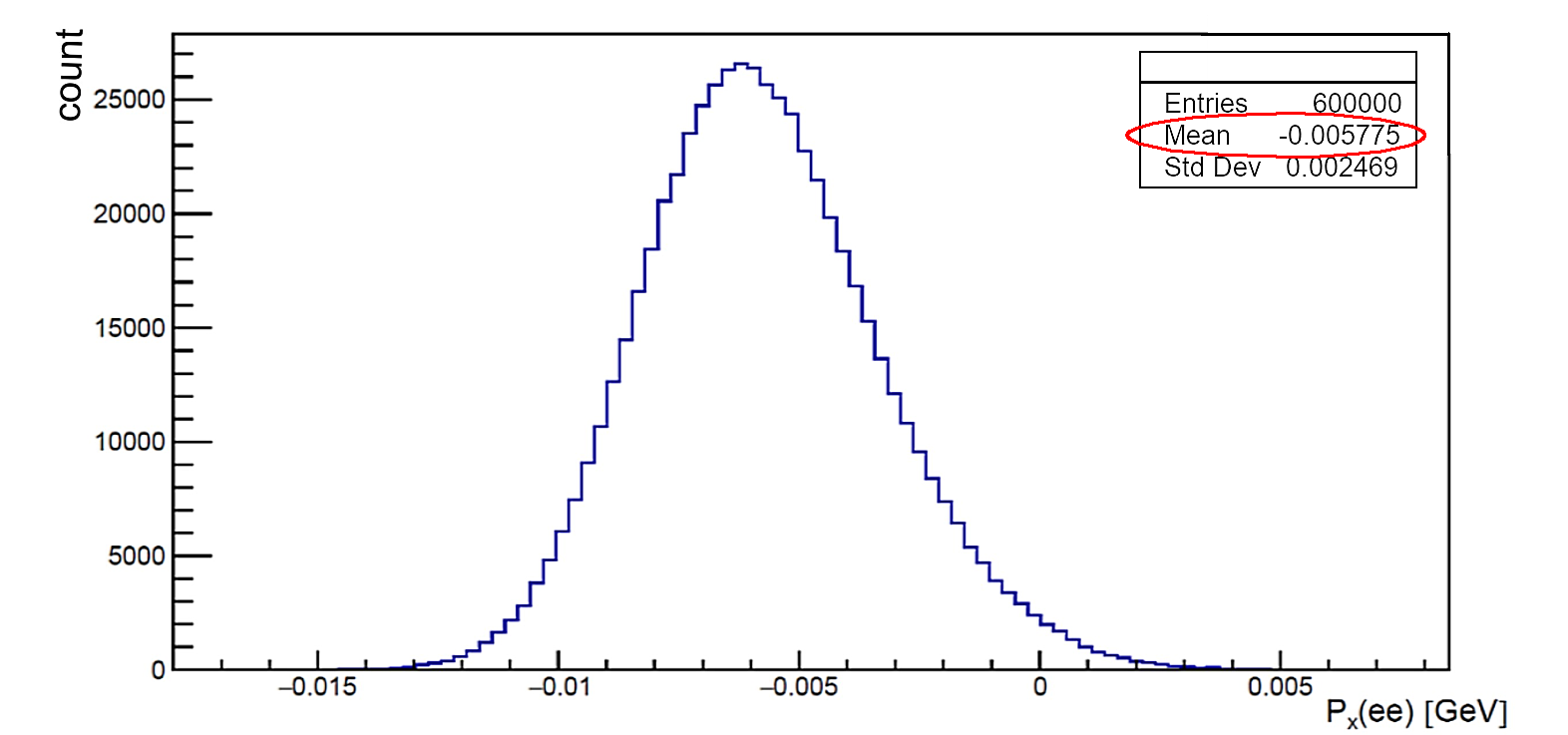}
\caption{Illustration of the change of momentum of colliding electron-positron system along the x-axis in the laboratory frame.}
\label{fig_2}
\end{figure}

\begin{figure}[!h]
\centering\includegraphics[width=.6\textwidth]{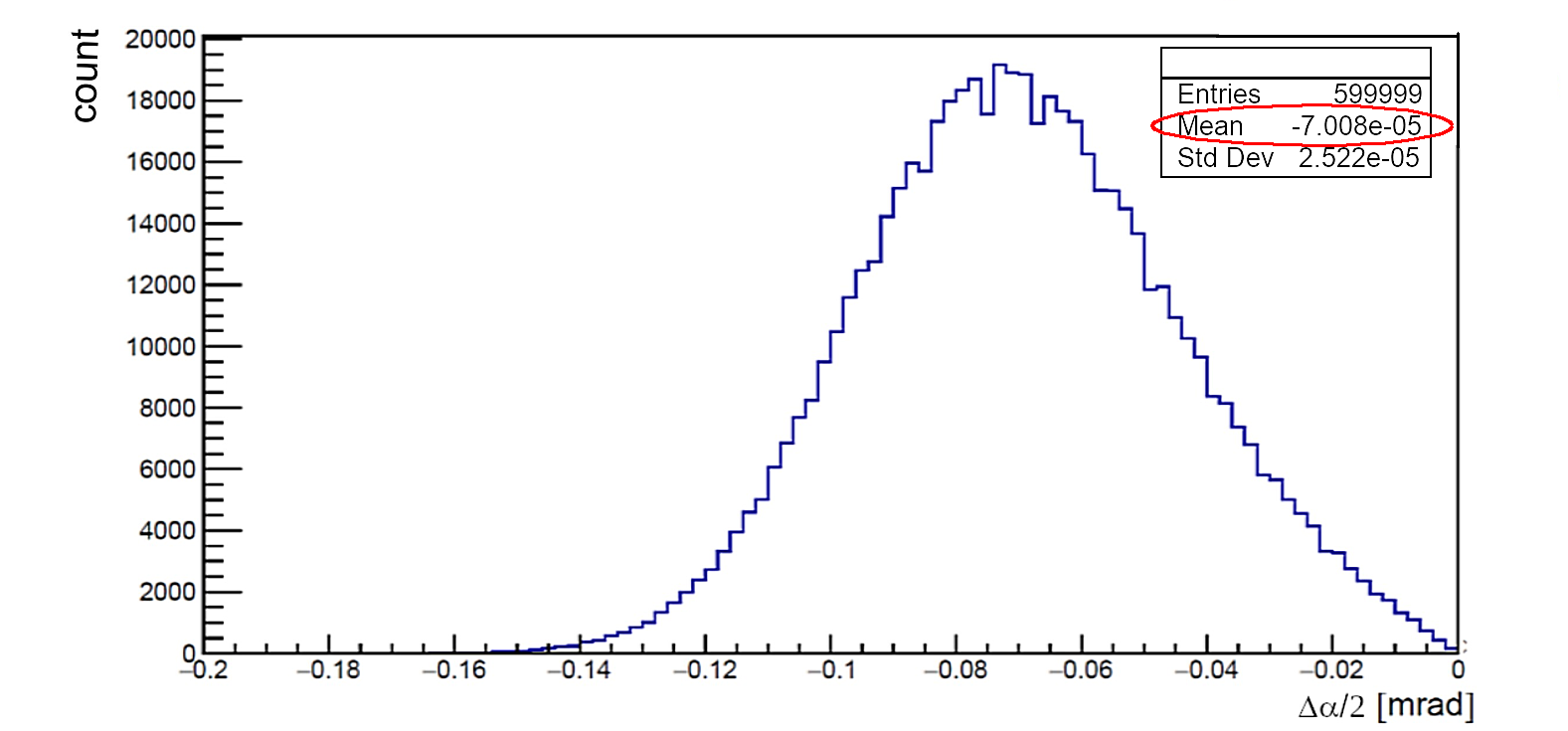}
\caption{Illustration of the effective reduction of the crossing angle per beam due to EMD1.}
\label{fig_3}
\end{figure}

\begin{figure}[!h]
\centering\includegraphics[width=.67\textwidth]{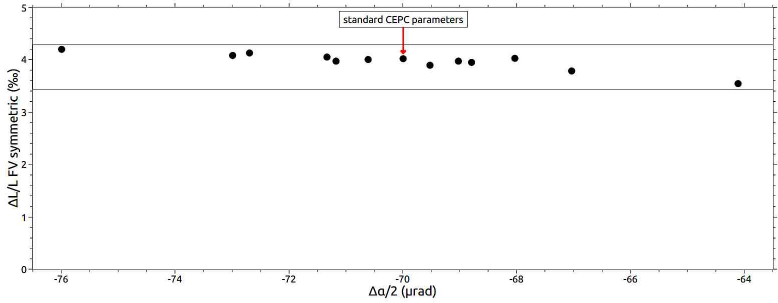}
\caption{Illustration of the EMD1 effect on the relative loss of integrated luminosity $\mathrm{\Delta\mathcal{L}_{int}/\mathcal{L}_{int}}$, versus reduction of the crossing angle $\mathrm{\Delta\alpha/2}$ introduced by this effect. Full lines indicate limits of $\mathrm{\Delta\mathcal{L}_{int}/\mathcal{L}_{int}}$ loss for $\pm$10\% variation of the nominal CEPC bunch parameters. The largest effect is induced by variation of the longitudinal bunch size.}
\label{fig_7}
\end{figure}

Due to EMD1, as can be seen from Fig. \ref{fig_1} and expected from equation \eqref{1.4}, the system of colliding particles is receiving a momentum kick in the negative direction of the laboratory x-axis. The mean of this kick is found to be $\mathrm{\sim5.8\: MeV}$, or $\mathrm{\sim2.9\: MeV}$ per initial state particle in average, assuming the nominal CEPC beam parameters from Table \ref{table1}). This is illustrated in Fig. \ref{fig_2}. Impact of beam parameter variations will be discussed later in this section, in the context of luminosity measurement (Fig. \ref{fig_7}). The EMD1 effect is restricted to the x-z plane and no bias is produced in the y-component of momentum of the colliding system. The crossing angle is in average reduced for $\mathrm{\sim140\: \mu rad}$, that is $\mathrm{70\: \mu rad}$ per beam. This is illustrated in Fig. \ref{fig_3}. In addition, the change of the crossing angle ($\mathrm{\Delta \alpha}$) of the interacting particles will modify components of their momenta and consequently their energies as: $\mathrm{\Delta E \sim E \cdot tg(\alpha/2) \cdot \Delta \alpha}$. Beam energy change is found to be $\mathrm{\Delta E \sim 52 \:keV}$ per CEPC beam in average.

The EMD1 effect will also produce consequent modifications of the Bhabha final state four-momenta. Using the BHLUMI V4.04 generator\footnote{While region of Bhabha polar angles bellow $\mathrm{\sim50\: \mu rad}$ is strongly dominated by t-channel photon exchange and regions above  $\mathrm{\sim100\: \mu rad}$ start to receive non-negligible contribution from s-channel Z exchange and $\mathrm{Z-\gamma}$ interference, between 50 and 80 mrad is the overlap region where the impact of difference between BHLUMI and BHWIDE \cite{BHWIDE} as Bhabha generation tools is at or below the required sensitivity limit for the $\mathrm{\mathcal{L}_{int}}$ measurement (in the context of observables relevant for discussion in this paper).} \cite{BHLUMI}, we have produced $\mathrm{\sim 6 \cdot 10^5}$ SABS events that we have associated with the initial states modified by EMD1 in GuinePig C++ V.1.2.2. Association here means modification of Bhabha electron and positron four-momenta through the boost to the reference frame of colliding particles and a consequent rotation forcing them back to back, performed on event-by-event basis. In this way a given initial system of colliding particles is becoming a center-of-mass frame of the considered final state electron and positron. Assumed position of the luminometer along the s-axis is preserving the geometry of head-on collisions, so no additional rotation due to the crossing-angle is performed.

\begin{figure}[!h]
\centering\includegraphics[width=.55\textwidth]{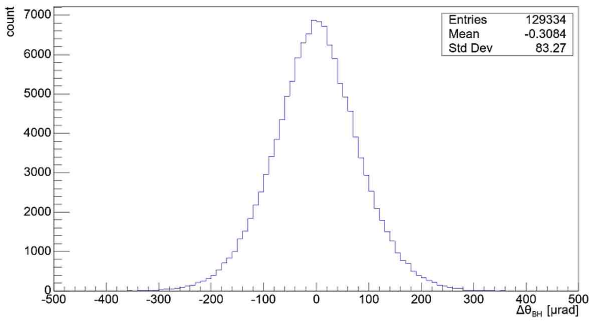}
\caption{Illustration of the smearing in polar angles ($\mathrm{\Delta \theta_{BH}}$) of the Bhabha final states due to EMD1, without biasing the mean.}
\label{fig_4}
\end{figure}

\begin{figure}[!h]
\centering\includegraphics[width=.60\textwidth]{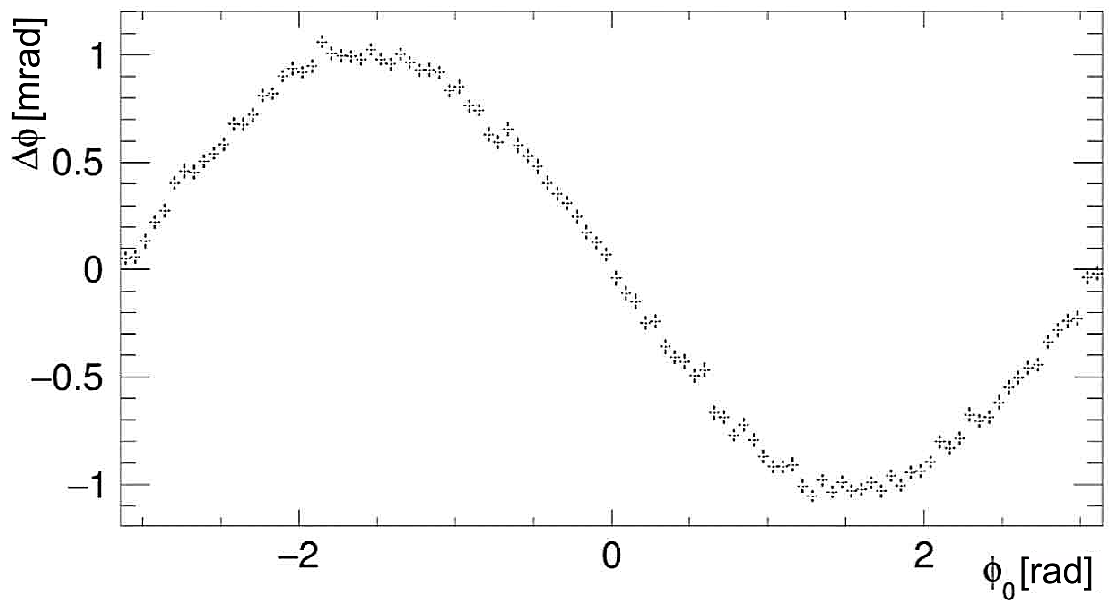}
\caption{Deviation of Bhabha azimuthal angles ($\Delta \phi$), due to EMD1, as a function of Bhabha final states azimuthal production angle $\phi_0$.}
\label{fig_5}
\end{figure}

The EMD1 effect results in smearing of the polar angle of the final state particles $\mathrm{\Delta \theta_{BH}}$ with RMS of $\mathrm{\sim 83 \: \mu rad}$ and without biasing the mean, what is illustrated in Fig. \ref{fig_4}. This can be intuitively understand in the following way: Suppose the incoming beams acquire small transverse angles ($\mathrm{\pm \Delta \alpha}$), the collision axis changes by roughly $\mathrm{\Delta \alpha}$, but the outgoing leptons are produced symmetrically around that axis. When transformed back to the detector frame some events shift to slightly larger polar angles, some to slightly smaller angles. To first order these shifts cancel statistically, producing no bias of the polar angle while distribution itself broadens. Due to the steep dependance of the Bhabha cross-section versus the polar angle ($\frac{d\sigma}{d\theta} \sim \theta^{-3}$), broadening near a sharp acceptance edge can still change event counts.  Maximal $\mathrm{\Delta \theta_{BH}}$ occurs for Bhabha events emitted along the x-axis. Deviation in azimuthal angles of the Bhabha final states are maximal ($\mathrm{\Delta \phi \sim 1 \: mrad}$) for SABS events emitted along y-axis, what is just a geometrical consequence of the initial state momentum shift along the x-axis and it is irrelevant for the luminosity measurement (unless event selection explicitly relies on Bhabha azimuthal position). The latter is illustrated in Fig. \ref{fig_5}. Although the EMD1 does not bias polar angles of final state electrons and positrons, it will result in modification of the angle between them, loss of collinearity and loss of count at the smallest polar angles.


\begin{figure}[!h]
\centering\includegraphics[width=.59\textwidth]{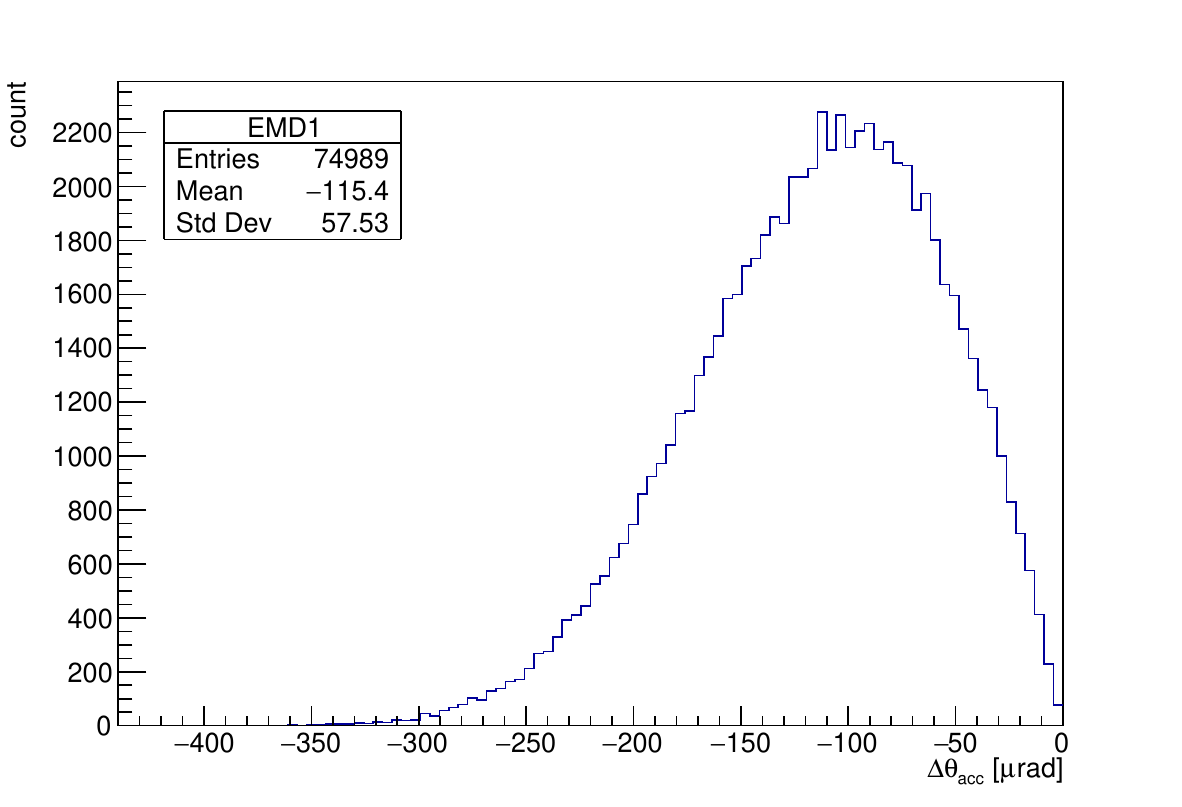}
\caption{Change in collinearity between final state electrons and positrons due to the EMD1 effect.}
\label{fig_6}
\end{figure}

This change of angle between final state electrons and positrons ($\mathrm{\Delta \theta_{acc}}$) is in average $\mathrm{ 115 \: \mu rad}$ (Fig. \ref{fig_6}) and it will affect SABS count in the luminometer in particular at the inner aperture where the effect is also strongest due to vicinity of opposite-charge bunches. The relative loss of count is found to be $\mathrm{\sim 4 \cdot 10^{-3}}$ what is $\mathrm{\sim 40}$ times larger than the integrated luminosity precision goal of $\mathrm{10^{-4}}$ in relative uncertainty. This loss can be taken as correction to the measured integrated luminosity, in particular if the crossing angle can be precisely measured in short intervals of time. This possibility is discussed in Section \ref{sec:sec5}. Even if the correction of count due to modified crossing angle is known, beam parameters may vary due to various effects including beam-beam interactions (up to 5\% at LEP \cite{Burkhardt}). Considering $\mathrm{\pm}$ 10\% variations of the nominal CEPC bunch parameters (Table \ref{table1}) as conservative,  changes of SABS count are given in Fig. \ref{fig_7}. Relative deviations of Bhabha count due to beam parameters variations are not larger than ($\mathrm{-5 \cdot 10^{-4}, \: +2 \cdot 10^{-4}}$) as indicated with the full lines in Fig. \ref{fig_7}. This means that once the correction to the Bhabha count is established (crossing angle is measured),  deviations induced by the bunch parameters variations up to 10\% from the nominal values will contribute to the relative systematic uncertainty of the integrated luminosity at most as $\mathrm{5 \cdot 10^{-4}}$. The EMD1 effect is the most sensitive to variations of the bunch length ($\mathrm{\sigma_z}$), what is obvious from the fact that the electromagnetic deflection of the incoming particles is generated during the overlap process itself, and the overlap geometry is controlled very strongly by the bunch longitudinal size. Does this conflict equation \eqref{1.5}? Actually, no, because CEPC operates in the large Piwinski angle regime ($\mathrm{\phi_P \gg 1}$) since $\mathrm{\sigma_x \ll \sigma_z}$. Effective overlap width than becomes:

\begin{equation}\label{1.6}
\sigma_x \cdot \sqrt{(1+\phi_{P}^2}) \approx \sqrt{(\sigma_{x}^2+\sigma_{z}^2 \cdot tg^2(\alpha /2))} \approx \sigma_z \cdot (\alpha /2),
\end{equation}

meaning that the overlap geometry is dominated by the bunch length, not by intrinsic transverse beam size.

In this paper, the estimates of Bhabha count in the luminometer are obtained by coincidental counting performed in the same polar angular regions in the left and right arms of the luminometer. If, however, one would count in a left-right asymmetric way, assuming (55-77) mrad counting angles in one half of the luminometer and (53-79) mrad in the other, subsequently applied to the left and right arm of the detector, the counting loss from EMD1 would be compensated to $\mathrm{\Delta \mathcal{L}_{int}/\mathcal{L}_{int} \sim 6 \cdot 10^{-5}}$ with the nominal CEPC beams, since this way of counting helps to cancel out the left-right asymmetry between the detector halves to which EMD1 is contributing\footnote{Beamstrahlung asymmetry, beam energy offsets, bunch phase offsets, detector alignment errors and interaction point offsets, together with other effects can contribute to the left-right asymmetry of the Bhabha experimental signature in the detector. Impact of the most relevant of these effects on integrated luminosity measurement at CEPC is discussed by the same authors in \cite{Mi}.}. However, for machines with the crossing-angle, this way of counting is meaningfully applicable only for a detector placed at the outgoing beams (s-axis).

\section{Electromagnetic deflection of the final state}
\label{sec:sec4}

Final state electron and positron produced in the Bhabha scattering will also be influenced by electromagnetic fields of the incoming bunches of opposite charges. This effect, labeled as EMD2, will rise with decreasing center-of-mass energy as it has been discussed in \cite{Cecile}, thus being of relevance at the $\mathrm{Z^0}$ resonance. While electromagnetic deflection of the initial state rotates (fluctuates) the collision frame, electromagnetic deflection of the final states directly bends the outgoing trajectories of Bhabha particles. After scattering, the outgoing electron (or positron) passes through the collective EM field of the outgoing opposite-charge bunch receiving the integrated Lorentz force as in equation \eqref{1.3}. Unlike the initial state deflection, this kick acts directly on the observed trajectory and not on the collision frame, changing the polar angle of final state particles $\mathrm{\Delta \theta_{BH}}$ as in equation \eqref{1.4}:

\begin{equation}\label{1.7}
\Delta \theta_{BH} \approx \frac{|\Delta \overrightarrow{p}_\perp|}{p}.
\end{equation}

The kick direction is not random. It has a preferred sign relative to detector acceptance, predominantly focusing outgoing particles toward the opposite beam axis. Therefore, EMD2 naturally produces a mean polar-angle bias that will be quantified further in simulation.

\begin{figure}[!h]
\centering\includegraphics[width=.59\textwidth]{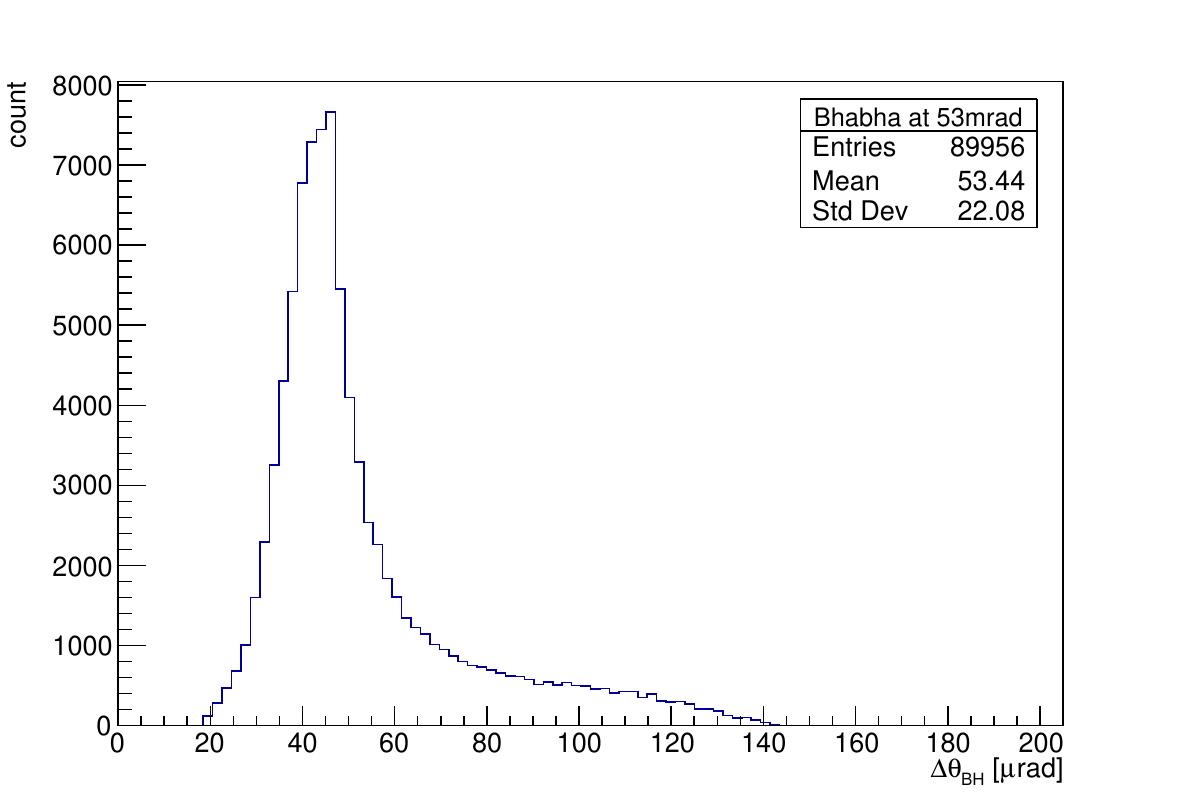}
\caption{Mean shift of polar angles ($\mathrm{\Delta \theta_{BH}}$) of Bhabha final state electrons at the luminometer inner edge (53 mrad), due to the EMD2 effect.}
\label{fig_8}
\end{figure}

\begin{figure}[!h]
\centering\includegraphics[width=.59\textwidth]{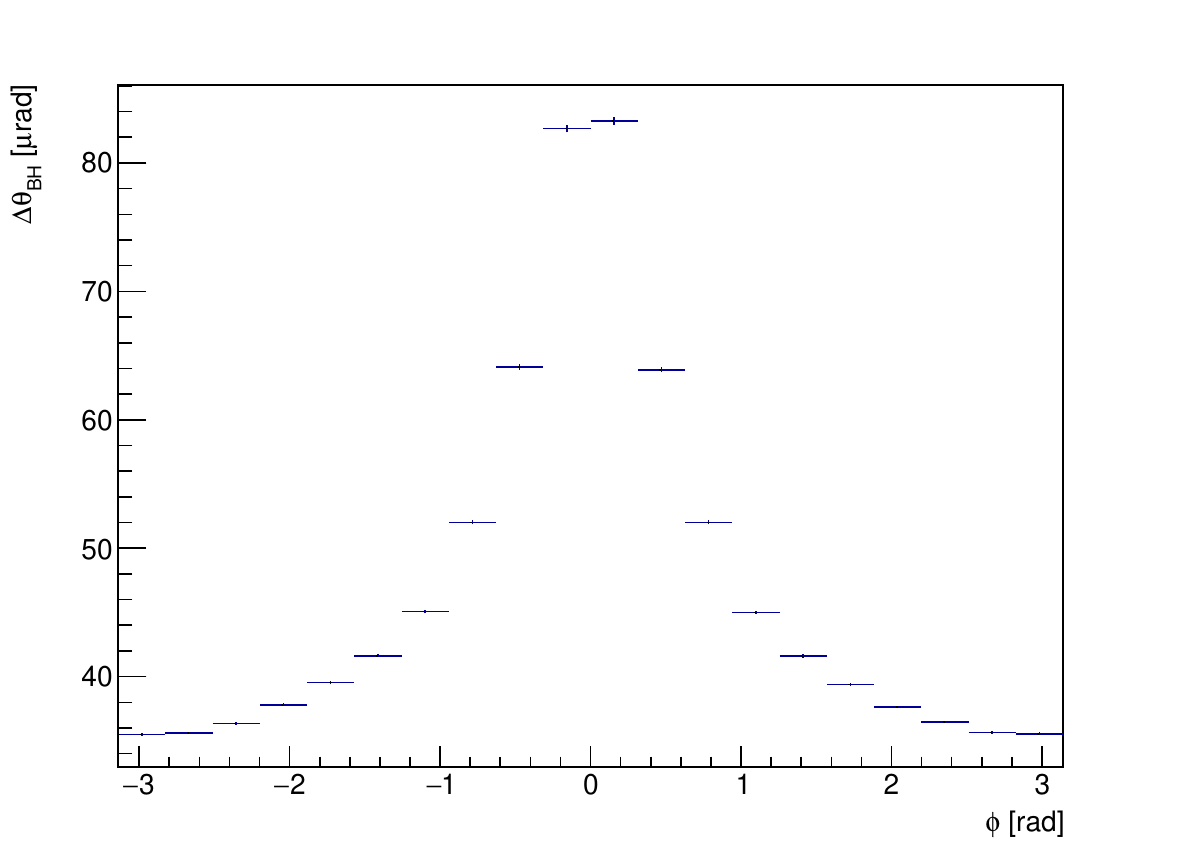}
\caption{Focusing the Bhabha final state electrons ($\mathrm{\Delta \theta_{BH}}$), induced by the EMD2 effect, as a function of the Bhabha production azimuthal angle.}
\label{fig_9}
\end{figure}

Bhabha electrons and positrons are generated slightly above the phase space of the luminometer fiducial volume (to allow particles to be focused towards luminometer fiducial volume) and have been tracked subsequently in the fields of the opposite-charge bunches using the Guinea Pig C++ V.1.2.2 software employed to simulate the EMD2 effect. In order to make a comparison with \cite{FCCee}, Bhabha events with the polar angle of 64 mrad are generated and mean deflection $\mathrm{\Delta \theta_{BH}}$ at CEPC is found to be 47.4 $\mu\mathrm{rad}$ compared to 41.2 $\mu\mathrm{rad}$ in \cite{FCCee}. However, inner aperture of the luminometer’s fiducial volume at CEPC is slightly lower – at 53 mrad, so the mean bias of the polar angle of particles emitted in this region is slightly larger - 53.4 $\mu\mathrm{rad}$. This is illustrated in Fig. \ref{fig_8}. Final state focusing angle $\mathrm{\Delta \theta_{BH}}$ caused by EMD2 also depends on the azimuthal angle of the final state particles, being the largest for particles emitted along the x-axis, similarly to the EMD1. For electrons, this is illustrated in Fig. \ref{fig_9}. This is fundamentally a geometry effect at a collider with a non-zero  crossing-angle. The strongest beam–beam field gradient experienced by the outgoing Bhabha scattering leptons is in the horizontal crossing plane. So particles emitted along the x-axis remain longest inside the strongest transverse field region of the outgoing opposite bunch and therefore accumulate the largest net kick.

\begin{figure}[!h]
\centering\includegraphics[width=.58\textwidth]{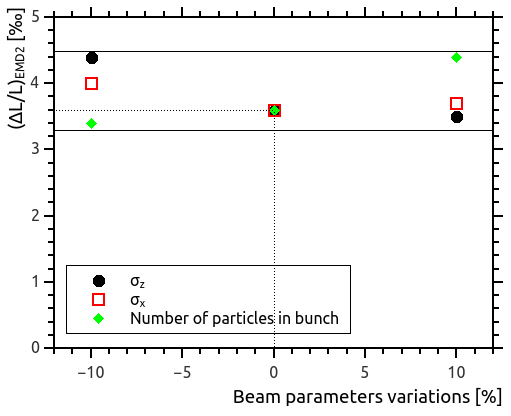}
\caption{Illustration of the EMD2 effect on the relative loss of integrated luminosity ($\mathrm{\Delta\mathcal{L}_{int}/\mathcal{L}_{int}}$) for variations up to $\pm$10\% of the nominal CEPC bunch parameters from Table \ref{table1}. Full lines indicate maximal variations of count. Initial and final state radiation are included.}
\label{fig_10}
\end{figure}

The loss of count in luminometer induced by the EMD2 is found to be $\sim 3.6 \cdot 10^{-3}$. Difference between simulation results obtained with BHLUMI and BHWIDE event generators are at the level of few times $\sim  10^{-4}$. Similarly to EMD1, final state electromagnetic deflection is relatively insensitive to $\mathrm{\pm}$10\% variation of the bunch parameters, where deviations of count are not larger than $(-0.2, +0.8)\,\text{\textperthousand}$ from the nominal beam parameters. This is illustrated in Fig. \ref{fig_10}. Like in the case of EMD1, the most relevant bunch parameter is the bunch length, in the large-Piwinski-angle regime. As given in equation \eqref{1.6}, the outgoing Bhabha scattering lepton traverses the opposite outgoing bunch field with the overlap geometry governed by $\mathrm{\sigma_Z}$ (and the crossing angle), followed closely by bunch population N.

\begin{figure}[!h]
\centering\includegraphics[width=.62\textwidth]{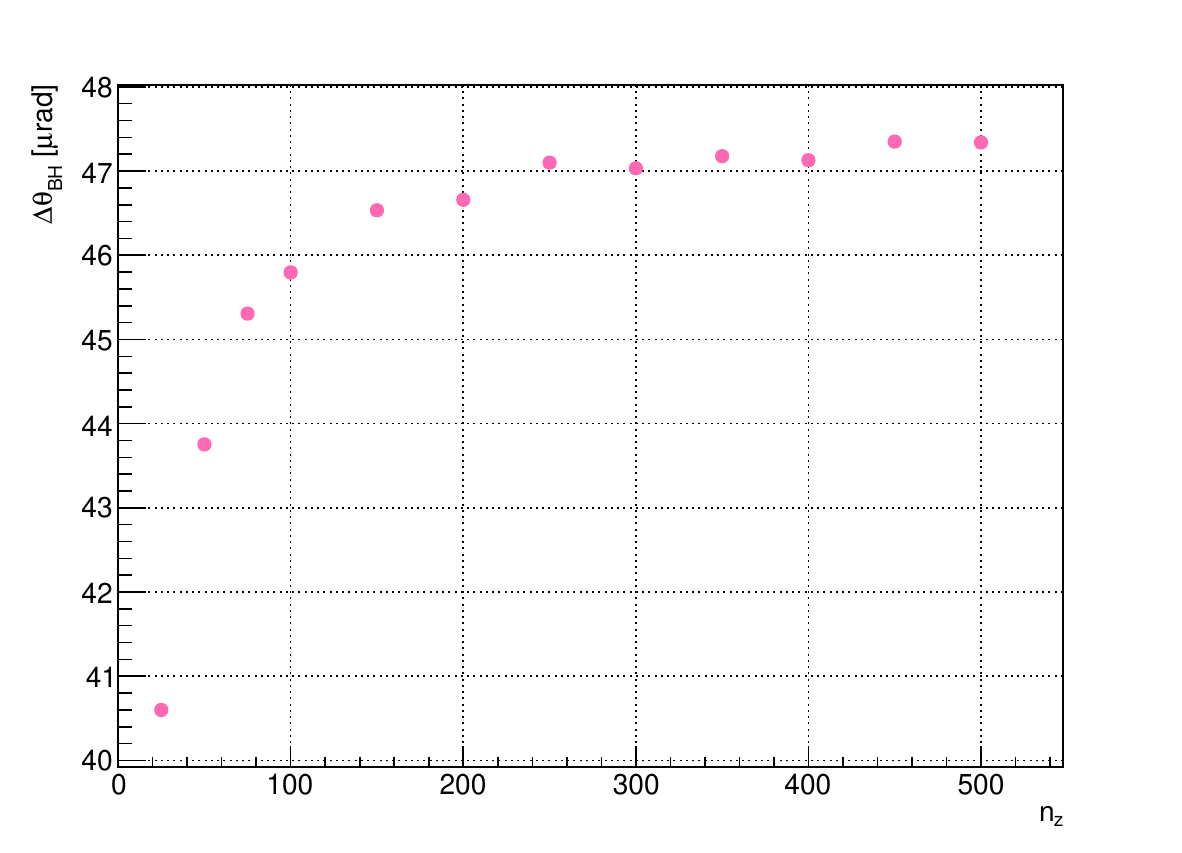}
\caption{Size of the Bhabha focusing ($\mathrm{\Delta \theta_{BH}}$) by EMD2, for events in the luminometer fiducial volume, with respect to the number of simulated bunch slices in the longitudinal direction ($\mathrm{n_z}$). No radiative effects are included.}
\label{fig_11}
\end{figure}

This effect is in principal sensitive to the settings of the simulation (as also shown in \cite{FCCee}), so in Fig. \ref{fig_11} we confirm stability of the result from Fig.  \ref{fig_8} with respect to the number of slices of a CEPC bunch in the longitudinal direction. The result from Fig. \ref {fig_8} is obtained with 250 longitudinal slices, that is in the saturation region from Fig.  \ref{fig_11}. 

It is also interesting to discuss the size of luminosity (count) uncertainty due to EMD1 and EMD2. It has been already illustrated that EMD1 mostly produces broadening (smearing) of final states polar angles (Fig.  \ref{fig_4}), while EMD2 produces a mean angular bias (Fig.  \ref{fig_8}). EMD2 may be comparable, or even dominate  EMD1 in terms of luminosity corrections because first-order acceptance migration dominates over symmetric broadening.

\section{Discussion}
\label{sec:sec5}
\begin{figure}[!h]
\centering\includegraphics[width=.57\textwidth]{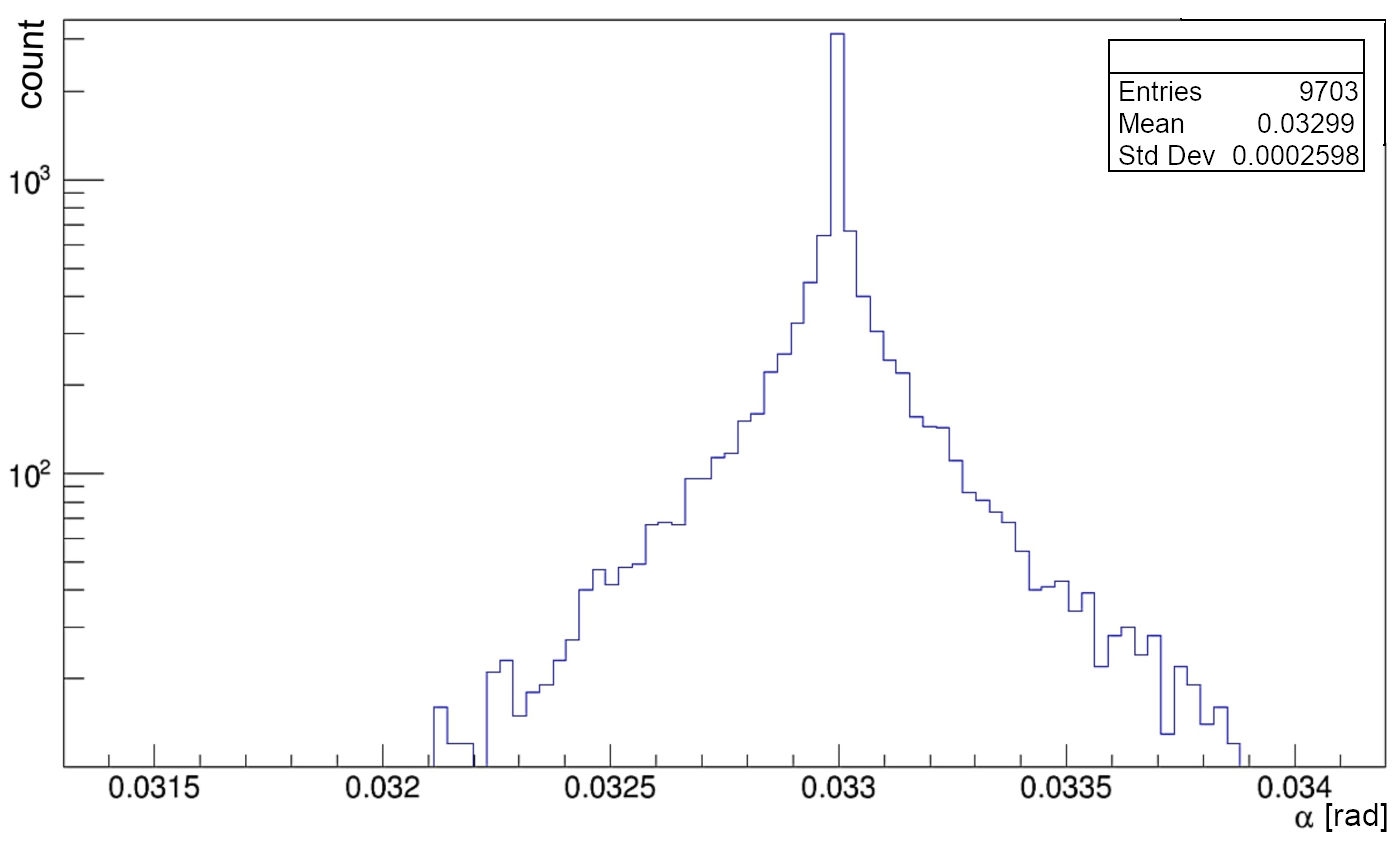}
\caption{Crossing angle determination from di-muon production at the $\mathrm{Z^0}$ pole, in $\sim$70 $\mathrm{pb^{-1}}$ of integrated luminosity (about 10 minutes).}
\label{fig_12}
\end{figure}

The correction to the loss of count induced by the EMD1 (Fig. \ref{fig_7}) can be introduced by measuring the exact crossing angle in the short interval of time. Uncertainty of integrated luminosity will be driven by precision the crossing angle is measured. The crossing angle at CEPC can be precisely determined from the central processes like the s-channel di-muon production. Kinematics of di-muon final states will be precisely measured in the central tracker where momenta of charged high energy leptons should be measured with the resolution of $\Delta p_t / p_t^2 \sim 10^{-5}\ \mathrm{GeV}^{-1}$. Considering  $\mathrm{\sim 10^4}$ di-muon events generated at the $\mathrm{Z^0}$ pole with WHIZARD V2.8.3 event generator\footnote{WHIZARD is a multi-purpose Monte Carlo event generator designed for simulation of high-energy particle collisions at lepton and hadron colliders based on tree-level determination of matrix elements for multi-particle final states. It includes interfaces for parton showering, ISR and beamstrahlung via external file.} \cite{WHIZARD} and assuming muons’ momenta smearing of order of $\mathrm{\sim 10^{-5} \: GeV^{-1}}$ to emulate detector resolution, the crossing angle can be measured with the standard error of $\mathrm{\sim 260 \: mrad/\sqrt{N_{\mu \mu}}}$ in about 10 minutes of the CEPC run at the $\mathrm{Z^0}$ pole. This is illustrated in Fig. \ref{fig_12}. In other words, crossing angle precision of order of $\mathrm{\sim 1 \: \mu rad}$ can be achieved with as little as $\mathrm{70 \: pb^{-1}}$ of integrated luminosity. While the EMD1 impact can be determined by crossing angle measurement, it is not yet clear, in terms of a full methodology confirmed in simulation, how to correct for EMD2 luminosity losses at CEPC. There are ongoing studies by the same authors to establish experimental correction of the luminosity losses induced by the EMD2 effect on the basis of probability regression using machine learning methods. 

We also discuss impact of radiative processes like initial state radiation (ISR), final state radiation (FSR) and beamstrahlung (BS) that will disturb four-momenta of LABS final states leading to counting losses at the luminometer’s edges. In principal, ISR emission will result in softer LABS final states that will be more strongly focused by the EMD2. Momentum loss induced by FSR should be recovered by the clustering algorithm since the FSR photons will be emitted in a narrow cone around the LABS final states, yet it will also reduce the Bhabha energy. Lower-energy outgoing leptons spend longer time inside the bunch field, receive larger angular perturbations and migrate more strongly across luminometer cuts. Impact of initial and final state radiation and beamstrahlung on Bhabha polar angle bias $\mathrm{\Delta \theta_{BH}}$ and relative luminosity losses $\mathrm{\Delta\mathcal{L}_{int}/\mathcal{L}_{int}}$ induced by EMD2, are given in Table \ref{table2}.


\begin{table}[!h]
\caption{Impact of radiative effects on Bhabha focusing and luminosity losses induced by EMD2, in the luminometer fiducial volume.}
\label{table2}
\centering
\begin{tabular}{|c|c|c|}
\hline
  & $\mathrm{\Delta\mathcal{L}_{int}/\mathcal{L}_{int}}$ & $\mathrm{\Delta \theta_{BH}\:  [\mu rad]}$ \\ 
\hline
No effects & 3.4 \textperthousand &  47.1 \\
\hline
ISR+FSR & 3.6 \textperthousand&  52.6\\
\hline
ISR+FSR+BS & 3.8 \textperthousand &  52.9 \\
\hline
\end{tabular}
\end{table}

Initial state radiation dominates FSR (second row in Table \ref{table2}), since ISR creates asymmetric low-energy tails, boosting effective collision frames and consequently producing acollinear configurations. Inclusion of beamstrahlung into Guinea Pig simulation on top of ISR/FSR will generally enhance the same mechanisms described above, but in circular colliders this effect is typically moderate rather than dramatic like at linear $\mathrm{e^+}$$\mathrm{e^-}$ colliders. Unlike ISR that is pure QED radiative process, beamstrahlung depends on machine parameters like bunch density, beam sizes and crossing angle being intentionally kept relatively small at circular colliders. For these reasons effect of BS is comparable to FSR and smaller than ISR, as illustrated in the third raw of Table \ref{table2}. However, it is worth noting that a fraction of luminosity loss can arise not only due to lower Bhabha momentum, but indirectly via changed beam-beam overlap because beamstrahlung may also broaden the (initial state) energy spectrum, modify effective crossing geometry and alter bunch focusing dynamically. Also, beam profile in a realistic experimental run will be non-Gaussian. Real bunches may have transverse tails, core-halo structure, asymmetries, bunch-to-bunch fluctuations, etc. Since beam-beam force depends on local density gradient, dense core regions will enhance focusing, while tails will weaken average focusing and asymmetries will create nonuniform kicks. However, luminosity will remain bunch core dominated and the effect of non-Guassian beam profile should not be larger than a few times $\mathrm{10^{-4}}$ on the Bhabha count, similar to radiative effects\footnote{This is also a sensitivity limit in this study due to finite samples' sizes (processing time).}. Guinea Pig itself assumes Gaussian beam profile by default, requiring external files for more realistic beam description, so this effect was not simulated.

\section{Conclusion}
\label{sec:sec6}

This is the first estimate of electromagnetic deflection effects at the CEPC $\mathrm{Z^0}$ pole. The effects of electromagnetic deflection of initial (EMD1) and final Bhabha states (EMD2) are quantified in simulation with the nominal post-CDR CEPC beams. These effects contribute to the loss of collinearity of Bhabha events that will be coincidently counted in left and right arms of the luminometer. Both effects will be simultaneously present and if uncorrected they will cause the relative loss of count in the luminometer $\mathrm{\Delta \mathcal{L}_{int}/\mathcal{L}_{int} \lesssim 6 \cdot 10^{-3}}$. Electromagnetic deflection of initial electrons and positrons towards bunches of the opposite charges will lead to reduction of the crossing angle of $\sim$140 $\mu\mathrm{rad}$. Measurement of the crossing angle with $\mathrm{\sim \: \mu rad}$ precision will be possible with the central tracker reconstruction of di-muon production, in approximately ten minutes  of the CEPC run at the $\mathrm{Z^0}$ pole. Once the crossing angle is corrected, bunch parameters variations within $\pm$10\% from the nominal values will produce variations of Bhabha count not larger than $5 \cdot 10^{-4}$ that will translate into the systematic uncertainty of the integrated luminosity. Luminosity bias induced by electromagnetic deflection of the final state is comparable in size to EMD1, as both effects originate from the same intense beam–beam fields and therefore generate angular perturbations of similar intrinsic scale. Although EMD1 mainly broadens the effective collision-frame orientation while EMD2 directly shifts the outgoing Bhabha angles, the steep small-angle Bhabha scattering acceptance makes both mechanisms capable of inducing luminosity biases at the $\mathrm{\sim 4 \cdot 10^{-3}}$ level. Variations in bunch sizes and population not larger than  $\pm$10\% introduce relative uncertainty of the Bhabha count due to EMD2 variations not larger than $\mathrm{8\cdot 10^{-4}}$. These estimates are conservative since beam parameters variations at LEP were typically at  $\sim$ 5 \% level. Radiative effects typically contribute to the integrated luminosity uncerainty as few times $\mathrm{10^{-4}}$. There is still no fully establish experimental method to correct for EMD2 at the CEPC $\mathrm{Z^0}$ pole, although the authors are carrying out a study based on machine learning to correct for this effect.

\section*{Acknowledgment}
This work is realized in the framework of MoU signed between the Vinča Institute of Nuclear Sciences – National Institute of the Republic of Serbia and the Institute of High Energy Physics (IHEP Beijing) of the Chinese Academy, on participation in the CEPC project.


%

\vspace{0.2cm}
\noindent


\let\doi\relax


\end{document}